\documentclass{ws-p9-75x6-50}

\usepackage{cite}

\def\be{\begin{equation}}
\def\ee{\end{equation}}

\begin{document}

\title{DC Transformer and DC Josephson(-like) Effects in Quantum Hall Bilayers}

\author{S. M. Girvin}

\address{
Yale University\\
Physics Department\\
Sloane Physics Lab\\
PO Box 208120\\
New Haven, CT 06520-8120 USA\\
 steven.girvin@yale.edu}




\maketitle

\abstracts{In the early days of superconductivity, Ivar Giaver
discovered that it was possible to make a novel DC transformer by
using one superconductor to drag vortices through another. An
analogous effect was predicted to exist in quantum Hall bilayers
and has recently been discovered experimentally by Eisenstein's
group at Caltech. Similarly, new experiments from the Caltech
group have demonstrated the existence of a Josephson-like
`supercurrent' branch for electrons coherently tunnelling between
the two layers.\\
\\
}

\section{Introduction}
The various quantum Hall effects are among the most remarkable
many-body phenomena discovered in the second half of the
twentieth century.
\cite{prangegirvin,tapash,perspectivesbook,leshouches} The
fractional effect has yielded fractional charge, spin and
statistics, as well as unprecedented order parameters.
\cite{ODLRO}  There are beautiful connections with a variety of
different topological and conformal field theories of interest in
nuclear and high energy physics.

The quantum Hall effect (QHE) takes place in a two-dimensional
electron gas formed in a quantum well in a semiconductor host
material and subjected to a very high magnetic field.  In essence
it is a result of a commensuration between the number of
electrons, $N$, and the number of flux quanta, $N_\Phi$, in the
applied magnetic field.  The electrons condense into distinct and
highly non-trivial ground states (`vacua') formed at each
rational fractional value of the filling factor $\nu \equiv
N/N_\Phi$.

The essential feature of (most) of these exotic states is the
existence of an excitation gap.  The electron fluid is {\em
incompressible} and flows rigidly past obstacles (impurities in
the sample) with no dissipation.
A weak external electric field
will cause the fluid to move, but the excitation gap prevents the
fluid from absorbing any energy from the electric field. Hence
the current flow must be exactly at right angles to the field
and the conductivity tensor takes the remarkable universal form
\be\sigma^{xx} = \sigma^{yy} = 0;\,\,\,\,\,
\sigma^{xy} = -\sigma^{yx} = \nu\frac{e^2}{h}.
\ee
Ironically, this ideal behavior occurs because of imperfections
and disorder in the samples which localize topological defects
(vortices) whose motion would otherwise dissipate energy.  In a
two-dimensional superconductor, such vortices undergo a
confinement phase transition at the Kosterlitz-Thouless
temperature and dissipation ceases.  In most cases in the QHE, an
analog of the Anderson-Higgs mechanism  causes the vortices to be
deconfined \cite{ODLRO} so that dissipation is strictly zero only
at zero temperature.  In practice, values of
$\sigma^{xx}/\sigma^{xy}$ as small as $10^{-13}$ are not difficult
to obtain at dilution refrigerator temperatures.

Recent technological progress in molecular beam epitaxy
techniques has led to the ability to produce pairs of closely
spaced two-dimensional electron gases. Strong correlations
between the electrons in different layers lead to a great deal of
completely new physics involving spontaneous interlayer phase
coherence. \cite{fertig,wenzee,perspectivesbook,kyangprl,kmoonprb,%
kyangprb,Schlieman,ady1PRL,ady2PRL,leonleoPRL,fogler,chetan,yogesh2,%
yogesh,veillette,burkov} This is the first example of a QHE system
with a finite-temperature phase transition.  This transition is in
fact a Kosterlitz-Thouless transition into a broken symmetry state
which is closely analogous to that of a 2D superfluid. Recent
remarkable tunnelling experiments by Eisenstein's group at Caltech
\cite{jpetunnel1,jpegoldstone} have observed something closely
akin to the Josephson effect in superconducting tunnel junctions
and have measured the dispersion of the superfluid Goldstone mode.
It was predicted some years ago that the broken symmetry state
should exhibit quantized drag \cite{kmoonprb,duan}, and the
Caltech group has recently observed this effect. \cite{mindy} This
experiment is analogous to the classic DC transformer effect
discovered by Ivar Giaver \cite{giaver} in which one
superconductor is used to drag vortices across another thereby
creating a 1:1 voltage transformer that works with DC rather than
AC currents.

\section{Giaver DC Flux Transformer}

Let us begin our analysis with a review of dissipation in a
superconducting film in the presence of a weak magnetic field.
The magnetic field will induce vortices so that the
superconducting order parameter is approximately given by
\be
\Psi(z) \sim \prod_j \frac{z-Z_j}{|z-Z_j|} = e^{i\varphi(x,y)}
\ee
where $z\equiv x+iy$ is a complex number representing the
position vector $(x,y)$ in the 2D film and $Z_j$ is the position
of the $j$th vortex in the same complex notation.  (I have
neglected the variation in the magnitude of the order parameter
in the core region of the vortices.)  The phase $\varphi$ of the
order parameter winds by $2\pi$ in circling a vortex
\be
\oint d\vec r\cdot\vec\nabla\varphi = 2\pi.
\ee
Under typical circumstances, the vortices in a superconductor
should be viewed as `heavy' classical objects with strong
dissipation due to the normal region in the core.  Hence these
objects are strongly coupled to the lattice and tend to remain at
rest.  However in the absence of actual pinning of the vortices to
disorder in the lattice, a bias current applied to the film will
cause the vortices to drift perpendicular to the current due to
the Magnus force.  For each vortex that drifts across the sample,
the phase difference along the direction of the current slips by
$2\pi$
\be
\frac{d(\varphi_2-\varphi_1)}{dt} = 2\pi \dot n_{\rm V}
\ee
where $\dot n_{\rm V}$ is the flux of vortices drifting across
the sample.  Using the Josephson relation
\be
\hbar \dot\varphi = 2eV
\ee
we conclude that the dissipative voltage drop along the current
direction is
\be
V_2 - V_1 = \frac{h}{2e} \dot n_{\rm V}.
\ee

The Giaver flux transformer consists of two superconducting films
separated by a thin insulating layer.  Provided that the magnetic
penetration depth is not too large, the magnetic field will be
inhomogeneously concentrated in the vicinity of the vortices.  As
a result there will be a magnetic coupling energy which will
prefer for the vortices in the two layers to be bound together.
When this binding force is strong enough, current applied to one
layer (the `primary') will drag the vortices together across both
the `primary' and the `secondary' layers so that
\be
\dot\varphi_{\rm secondary} = \dot\varphi_{\rm primary}
\ee
and the voltage drop will be identical in both layers.  Hence the
system acts as a precise 1:1 voltage transformer. \cite{giaver}

\section{Coulomb Drag DC Transformer}

In addition to this magnetic coupling mechanism, one might
imagine that Coulomb interactions between the electrons in the
two superconducting films might lead to a drag effect.  For layer
separation $d$,  the interlayer Coulomb interaction has strength
\be
U_{\rm inter} = \frac{2\pi}{q}e^{-qd}
\ee
for momentum transfer $\hbar q$ between the layers. However
because the layer spacing ($\sim 3-10$nm) is considerably larger
than the spacing between electrons within a layer, $k_{\rm F}d\gg
1$ and the interaction is negligible.   Each electron gas sees the
other as an essentially structureless continuum.

Coulomb interactions are however very important if we consider
drag in a pair of closely spaced two-dimensional electron gases
(2DEGs) created in semiconductor quantum wells. Here the layer
spacing is also on the scale of $10$nm, but the electron density
is vastly lower so that $k_{\rm F}d\sim 1$ and the interlayer
Coulomb interactions are comparable in strength to the intralayer
interactions. Consider first the force balance in a single layer.
Newton's law for the carrier drift velocity in the presence of an
electric field $E$ is
\be
\dot v = -\frac{e}{m} E - \left(\frac{1}{\tau_{\rm L}} +
\frac{1}{\tau_{\rm ee}} \right)v
\ee
where $1/\tau_{\rm L}$ is the relaxation rate due to collisions
with impurities in the lattice and $1/\tau_{\rm ee}$ is the
relaxation rate due to electron-electron collisions.  However due
to galilean invariance, electron-electron collisions conserve the
center of mass momentum of the colliding particles and hence can
{\em not} relax the current and so $1/\tau_{\rm ee}=0$. Assuming
steady state ($\dot v=0$) and relating the drift velocity to the
current density via $J=-nev$ we obtain the standard result:
\be
J = \frac{ne^2\tau_{\rm L}}{m} E.
\ee
Thus even though Coulomb interactions produce strong correlations
and collisions, their effect is invisible in ordinary transport
measurements.

Drag measurements do not suffer from this problem and are an
excellent probe of Coulomb interactions
\cite{gramila,zhangmacd}.  To see how this works, consider the
force balance in a pair of closely spaced layers denoted by
$\uparrow,\downarrow$:
\begin{eqnarray}
\dot v_\uparrow &=& -\frac{e}{m}E_\uparrow - \frac{1}{\tau_{\rm
L}}v_\uparrow - \frac{1}{\tau_{\rm
D}}\left(v_\uparrow-v_\downarrow\right)\nonumber\\
\dot v_\downarrow &=& -\frac{e}{m}E_\downarrow -
\frac{1}{\tau_{\rm L}}v_\downarrow + \frac{1}{\tau_{\rm
D}}\left(v_\uparrow-v_\downarrow\right).
\end{eqnarray}
Here $1/\tau_{\rm D}$ is the rate of momentum transfer between
the layers due to electron-electron interactions which attempt to
relax the two layers towards a common center of mass velocity.
Galilean invariance only applies to the sum of the velocities in
the two layers, not the difference, and hence this relaxation term
does not vanish.  Assuming steady state and requiring that there
be no current flowing in the secondary layer ($\downarrow$, say)
we can readily solve for the drag induced electric field
\be
E_\downarrow = -\rho_{\rm D} J_\uparrow
\label{eq:transresistance}
\ee
where $\rho_{\rm D}$ is the drag or transresistance given by
\be
\rho_{\rm D} = +\frac{1}{\frac{ne^2\tau_{\rm D}}{m}}.
\ee
Equivalently
\be
E_\downarrow = - \frac{\tau_{\rm D}^{-1}}{\tau_{\rm
L}^{-1}+\tau_{\rm D}^{-1}}E_\uparrow.
\ee
The minus sign indicates that the electric field induced in the
drag layer is opposite in direction to that in the drive layer. (I
have chosen the standard sign convention in
Eq.~(\ref{eq:transresistance}) which makes the transresistance
positive.)  The sign is readily understood from the fact that
collisions with the electrons in the drive layer tend to push
electrons in the passive layer to flow in the same direction as
those in the drive layer.  However because we are insisting on
zero current in the drag layer, charge builds up until the
opposing electric field stops the current flow.  The electric
field force on the drag layer is thus opposite to that in the
drive layer.

Despite the Coulomb interactions, if the density is not too low,
the electrons in the two layers constitute fermi liquids in which
Pauli blocking severely limits the phase space available for
collisions.  Only electrons that lie within $k_{\rm B}T$ of the
fermi energy can participate in collisions and so we anticipate
that
\be
\frac{1}{\tau_{\rm D}} \sim \frac{1}{\tau_0}\left(\frac{k_{\rm
B}T}{\epsilon_{\rm F}} \right)^2
\ee
where $\tau_0$ is some microscopic collision rate in the absence
of Pauli blocking.  Hence interlayer collisions become less and
less important as the temperature decreases and we expect the
drag resistance to be small and ultimately vanish
\cite{gramila,zhangmacd}
\be
\rho_{\rm D} \sim + T^2.
\ee
This is precisely what was observed by Gramila et al.
\cite{gramila} At $T=1$K, they observed a drag resistance of only
a few milli-ohms per square in a pair of 2DEGs.

\section{Quantum Hall Coherent States}

\begin{figure}[t]
 \centerline{
\epsfysize=3.00in \epsffile{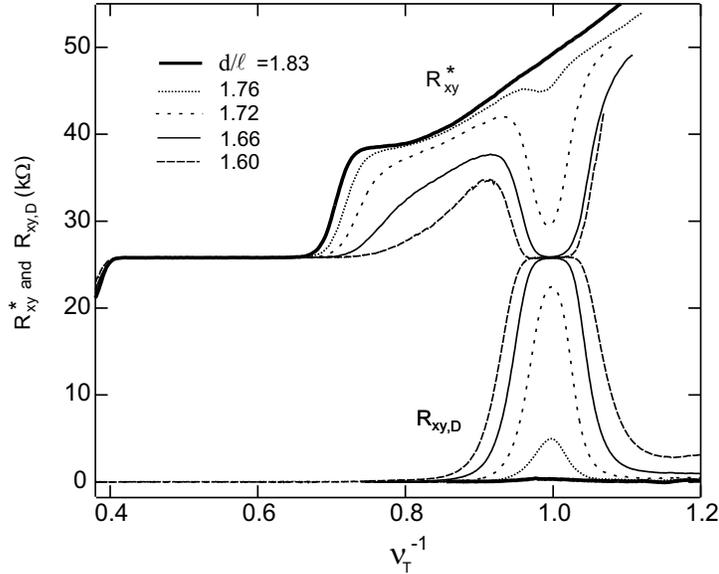}}
 \caption{Hall resistance (upper set of curves) and Hall drag
 resistance (lower set of curves) in a QHE bilayer system as a
 function of the inverse total filling factor
 $\nu_{\rm T}^{-1} = 1/(\nu_\uparrow + \nu_\downarrow)\propto B$.
 For small layer separation $d$ relative to the magnetic length
 $\ell$, there is a Hall plateau at
 $\nu_\uparrow=\nu_\downarrow=1/2$ in which the Hall field is
 identical in drive and drag layers even though no net current is
 flowing in the drag layer.
  After Kellogg et al. Ref.~\citelow{mindy}
 \label{fig:qhedrag} }
\end{figure}

Non-fermi liquids are of great current interest in the study of
strongly correlated systems.  It is easy to destroy 2DEG fermi
liquids by applying a strong external magnetic field which
quenches the kinetic energy and places the system in the quantum
Hall regime. \cite{leshouches}  As shown in
Fig.~(\ref{fig:qhedrag}), the drag in this case rises by some 6
orders of magnitude and for certain values of the magnetic field,
the transverse (Hall) drag resistance takes on a universal value
\be
\rho_{\rm D}^{xy} = - \frac{h}{e^2} \sim -25,813\Omega.
\ee
This is very different from the fermi liquid result because the
electric field is at right angles to the current, does not vanish
at low temperatures, and is in the {\em same}, not opposite,
direction in both layers.

We begin our analysis of the QHE regime with the simplest example
of the integer QHE in a single layer system of spinless electrons
at Landau level filling factor $\nu=1$. The strong magnetic field
quantizes the kinetic energy into discrete Landau levels
\cite{leshouches} separated in energy by the cyclotron energy
$\hbar \omega_{\rm c} \sim 100$K. Each level has  a {\em
macroscopic} degeneracy equal to $N_\Phi$. This degeneracy in the
kinetic energy means that interactions are enormously important
and have non-perturbative effects at fractional filling factors.
However for $\nu\equiv \frac{N}{N_\Phi}=1$, every state of the
lowest Landau level (LLL) is occupied and, since there is a large
kinetic energy gap to the next Landau level, interactions are
(relatively) unimportant.  It is this gap which makes the system
incompressible. Since the lowest LLL is completely full, the state
is a simple Slater determinant.
In first-quantized form the state is most easily expressed in the
symmetric gauge \cite{leshouches}
\be
\Psi(z_1,z_2,\ldots,z_N) = \prod_{i<j}^N (z_i - z_j) e^{-
\frac{1}{4}\sum_m |z_m|^2}
\label{eq:laughlin}
\ee
where $z_j \equiv (x_j + iy_j)/\ell$
is a dimensionless complex number representing the 2D position
vector of the $j$th particle in units of the magnetic length
$\ell$.  The vandermonde polynomial factor in this Laughlin state
is totally antisymmetric and is equivalent to a single Slater
determinant filling all the orbitals in the LLL.

The meaning of the Laughlin wave function can be seen by
comparison with the superconducting vortex problem.  In the QHE
regime vortices are not heavy classical objects with normal cores,
but rather highly quantum objects which attach themselves to the
electrons. Eq.~(\ref{eq:laughlin}) tells us that each electron
sees a complex zero of the wave function (i.e., a vortex) at the
position of each and every {\em other} electron.  Thus as current
flows through the device, exactly one vortex passes through for
each electron that passes through.  The vortex flux is therefore
simply related to the total current $I$
\be
\dot n_{\rm V} = \frac{I}{e}
\label{eq:nVdot}
\ee
and the Josephson relation (now for charge $e$, not $2e$) yields
the universal result
\be
V = \frac{\hbar}{e}\dot\varphi = \frac{\hbar}{e} 2\pi\dot n_{\rm
V}= \frac{h}{e^2} I.
\label{eq:rhoxy}
\ee
Since the vortex motion is now parallel rather than perpendicular
to the current, the (Hall) voltage drop is perpendicular to the
current and the flow is {\em dissipationless}.   Moving the
magnetic field away from the point which gives filling factor
$\nu$ exactly unity introduces extra vortices (or antivortices)
into the ground state, but in the presence of random disorder
these are pinned (just as in a disordered superconductor) and do
not affect the transport.  This is what allows the quantized
plateau to exist over a finite range of magnetic fields rather
than just at a single unique value of $B$.

\begin{figure}[ht]
 \centerline{
\epsfysize=3.00in \epsffile{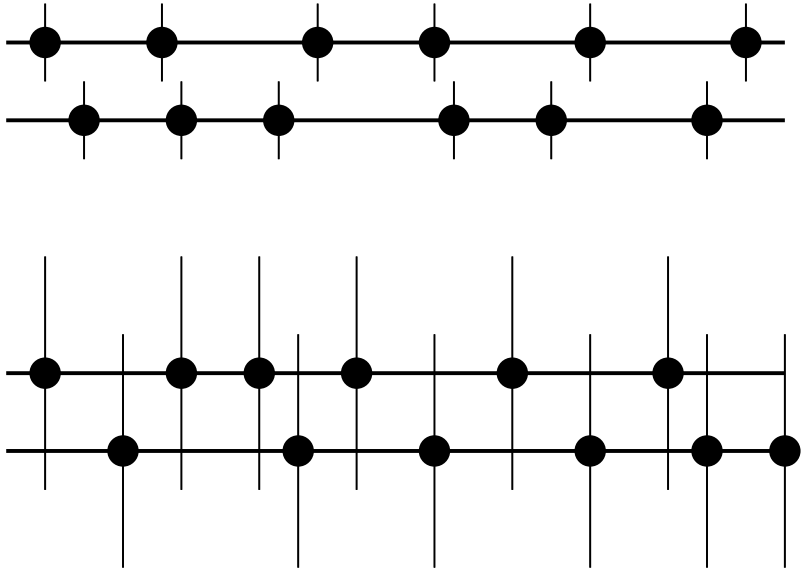}}
 \caption{
Cartoon of the bilayer quantum state.  Upper panel:  $\nu=1$ in
both layers. Electrons in the drag layer do not see the vortices
attached to the electrons in the drive layer.  Lower panel:  The
magnetic field is twice as large so that $\nu=1/2$ in each layer
and there are now twice as many vortices as electrons in each
layer. Each electron sees vortices attached to {\em all} electrons
including those in the other layer.
 \label{fig:cartoon} }
\end{figure}

We see this very wide plateau in the QHE drag data in the left
portion of Fig.~(\ref{fig:qhedrag}) corresponding to $\nu=1$ in
each layer.  In this regime the drag voltage vanishes
exponentially with temperature because each layer has an
excitation gap and so it is not possible at low temperatures to
have any excitations produced by interaction between the layers.
In the language of vortices we can understand this result using
the cartoon representation of the state shown in
Fig.~(\ref{fig:cartoon}).  At filling factor $\nu=1$ in each
layer ($\nu_{\rm T}=2$), there are exactly as many vortices in
each layer as there are electrons.  The quantum state satisfies
this condition by having each electron see only vortices attached
to electrons in the {\em same} layer.  Hence the electrons in one
layer do not see the vortices in the other layer.  (If they did,
then they would see a total of too many vortices.)  As a result,
when current flows in the drive layer, the electrons in the drag
layer do not see any vortices moving by and the drag voltage
vanishes.

At filling factor $\nu=1/2$ in each layer, the magnetic field is
twice as large and there are now twice as many vortices as
electrons in each layer.  The Coulomb energy favors the state
shown in the lower panel of Fig.~(\ref{fig:cartoon}) in which each
electron sees a vortex attached to {\em every} electron whether
they are in the same or different layers.  Because the vortices
are complex zeros of the wave function, the electrons strongly
avoid each other, independent of whether they are in the same or
different layers. This state turns out to be the {\em exact}
ground state for zero layer spacing and is a good approximation
for small $d/\ell$.

Because electrons see vortices from all the other electrons, a
current in the drive layer drags vortices through the lower layer
at a rate given by Eq.~(\ref{eq:nVdot}).  Again applying the
Josephson relation as in Eq.~({\ref{eq:rhoxy}) yields a universal
quantized Hall transresistance:
\be
\rho_{\rm D}^{xy} = -\frac{h}{e^2}.
\ee
The electric field is perpendicular to the current and has exactly
the {\em same} sign and magnitude as the field in the drive layer.
This prediction \cite{kmoonprb,duan,kyangdrag,kimdrag} is
beautifully and precisely verified in the recent experiment of
Kellogg et al.\ shown in the right portion of
Fig.~(\ref{fig:qhedrag}).

Because every electron sees a vortex attached to every other
electron, the microscopic wave function for this special state is
simply that given by Eq.~(\ref{eq:laughlin}).  We first wrote this
down above for a single layer, but it applies here because the
wave function is completely independent of which of the two layers
any electron is in!
\cite{helvphysacta,fertig,wenzee,kmoonprb,duan,kyangdrag,kimdrag}.
This is an explicit manifestation of the strange fact that, even
though tunnelling between the layers might be forbidden, quantum
mechanics still allows for the possibility of states in which we
are uncertain which layer the electrons are in.  It is very useful
to introduce a pseudospin 1/2 to represent the layer index,
$|\uparrow\rangle$ representing the electron being in the upper
layer and $|\downarrow\rangle$ corresponding to the lower layer.
[We assume that the real spin is frozen out by the applied
magnetic field.] In this language the microscopic wave function
becomes
\be
\Psi(z_1,z_2,\ldots,z_N) = \prod_{i<j}^N (z_i - z_j) e^{-
\frac{1}{4}\sum_m |z_m|^2} \left|
\rightarrow\rightarrow\rightarrow\rightarrow\rightarrow\ldots\rightarrow\right\rangle,
\label{eq:ferromagnet}
\ee
where the arrows represent the coherent spinors
\be
|\rightarrow\rangle = \frac{1}{\sqrt{2}}\left(|\uparrow\rangle +
e^{i\varphi}|\downarrow\rangle\right).
\ee
This corresponds to the pseudospin lying in the xy plane at an
angle $\varphi$ away from the x axis.  Because the spatial part of
the wave function is fully antisymmetric (which optimizes the
Coulomb energy), the pseudospin part must be fully symmetric
implying that this is a pseudospin ferromagnet.  Each electron is
in a coherent superposition of both layers. The microscopic wave
function in Eq.~(\ref{eq:laughlin}) actually corresponds to the
special case of $\varphi=0$.  However because tunnelling is
absent, there is a global symmetry (corresponding to the
conservation of $N_\uparrow - N_\downarrow$) which tells us that
there is actually an entire family of degenerate states with
different values of $\varphi$.  That is, there is a spontaneously
broken U(1) symmetry \cite{fertig,wenzee} in which the order
parameter
\be
\langle \sigma_x+i\sigma_y\rangle =
\langle\psi^\dagger_\uparrow\psi_\downarrow\rangle \sim
e^{i\varphi}
\label{eq:orderparameter}
\ee
condenses.

While the energy can not depend on the global value of $\varphi$
it can depend on spatial gradients
\be
H = \frac{1}{2}\rho_s \int d^2r |\nabla\varphi|^2
\ee
where the pseudospin stiffness $\rho_s$ represents the loss in
Coulomb exchange energy between the two layers when $\varphi$
varies with position.  The gradient energy is stored in a
`supercurrent'
\be
\vec J_- = \rho_s \vec\nabla \varphi.
\ee
Because the `charge' conjugate to $\varphi$ is $\sigma_z\sim
n_\uparrow - n_\downarrow$, this supercurrent is oppositely
directed in the two layers.  With this knowledge in hand we can
reanalyze the drag experiment in terms of the symmetric and
antisymmetric currents
\be
\vec J_\pm = \vec J_\uparrow \pm \vec J_\downarrow.
\ee
The symmetric channel transport is that of a $\nu=1$ quantized
Hall plateau with $\rho_+^{xy} = h/e^2$.  The antisymmetric
channel transport is that of a superfluid with $\sigma_-^{xx} =
\infty$.  In the drag experiment there is only current in the
drive layer so that
\be
\vec J_+ = \vec J_- = \vec J_\uparrow.
\ee
The symmetric current produces a quantized Hall electric field
which is identical in both layers and the superfluidity means
that the antisymmetric current produces no field at all.  This is
precisely the effect observed in the drag experiment by Kellogg
et al. and provides strong evidence for the existence of
superfluidity in this special interlayer phase coherent state.

Further strong evidence for phase coherence was discovered in a
remarkable tunnelling experiment by Spielman et
al.~\cite{jpetunnel1} in samples in which an extremely weak
tunnelling amplitude between the layers was used as a sensitive
probe. Because the order parameter in
Eq.~(\ref{eq:orderparameter}) is the tunnelling operator itself,
it is possible to tunnel an electron from one layer to the other
and still be in the same quantum state!  This paradox arises from
the fact that we were uncertain which layer the electron was in
originally.  If the quantum state is unchanged then energy is
conserved only at zero bias voltage. The Caltech group indeed
observed an enormous and extremely narrow (half width $<5\mu$V)
zero bias anomaly in the tunnelling as shown in
Fig.~(\ref{fig:tunnel1}). Unlike the true Josephson effect the
dissipation is not infinitesimal on the supercurrent branch.
Various proposals involving a finite phase coherence time have
been made to explain the finite height and width of the
differential conductance peak.
\cite{ady2PRL,leonleoPRL,fogler,yogesh2} but this is a question
which is still poorly understood and is a subject of current
study.

If the layer separation is increased, the system undergoes a
quantum phase transition to a quantum disordered state in which
the zero bias anomaly disappears and is replaced by a Coulomb
pseudogap as shown in the right hand panel of
Fig.~(\ref{fig:tunnel1}).

\begin{figure}[ht]
 \centerline{
\epsfysize=3.00in \epsffile{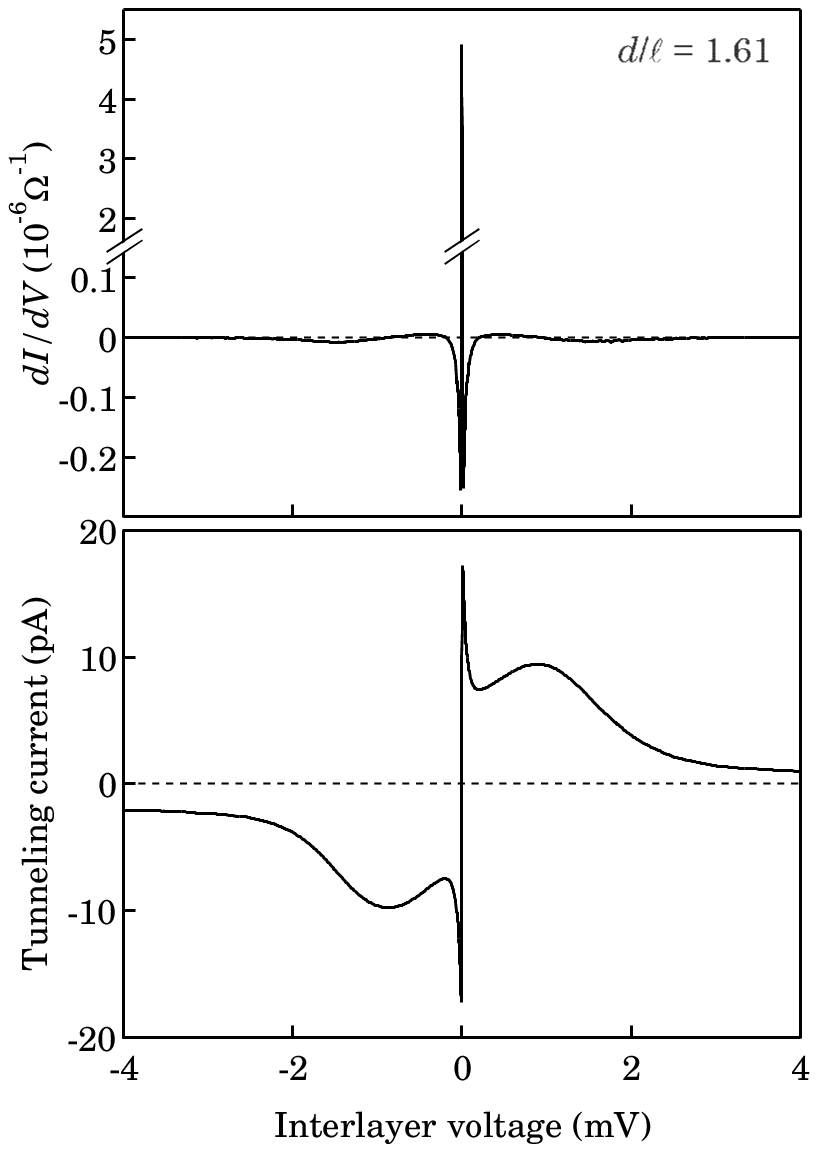}
 \epsfysize=2.95in \epsffile{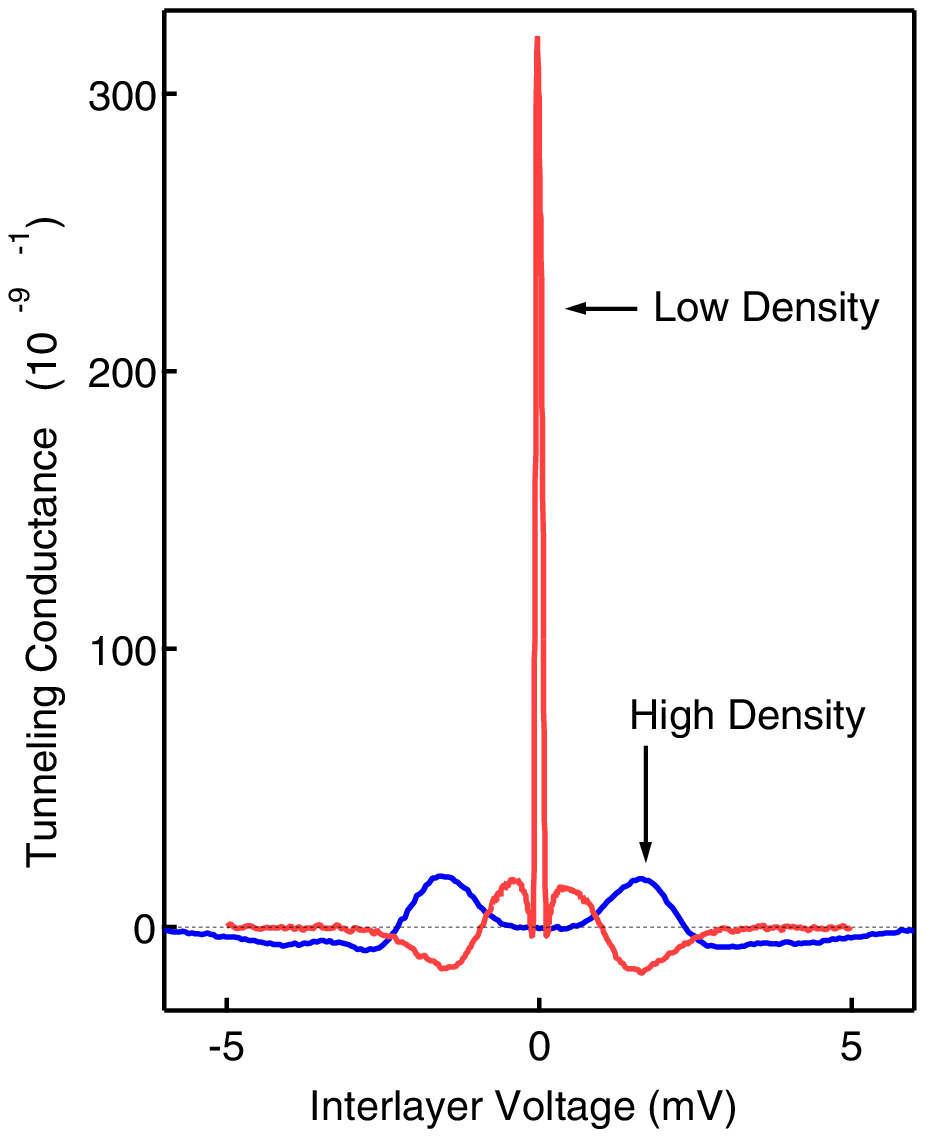} }
 \caption{Upper left panel:  Differential conductance of a QHE bilayer
 system in the phase coherent state at filling factor $\nu=1$ and
 layer spacing $d/\ell=1.61$.  The central peak is remarkably
 narrow with a HWHM of only about $6\mu$eV.  Lower left panel:  $IV$
 curve showing the nearly vertical `supercurrent' branch and a
 remnant of the Coulomb gap feature at larger voltages.
After Spielman et al. Ref.~\citelow{jpegoldstone}.
Right panel:
 Differential conductance at low density (small $d/\ell$) in
 the phase coherent state and high density (large $d/\ell$) where the layers
 are uncorrelated.  In the latter case the tunnel current vanishes at
 small voltages due to the Coulomb gap. There is a peak in the current at a
 voltage corresponding to the scale of the Coulomb interactions in the
 system. Figure courtesy of J.~P.~Eisenstein. \label{fig:tunnel1} }
\end{figure}

\begin{figure}[ht]
\centerline{ \epsfxsize=2.5in \epsffile{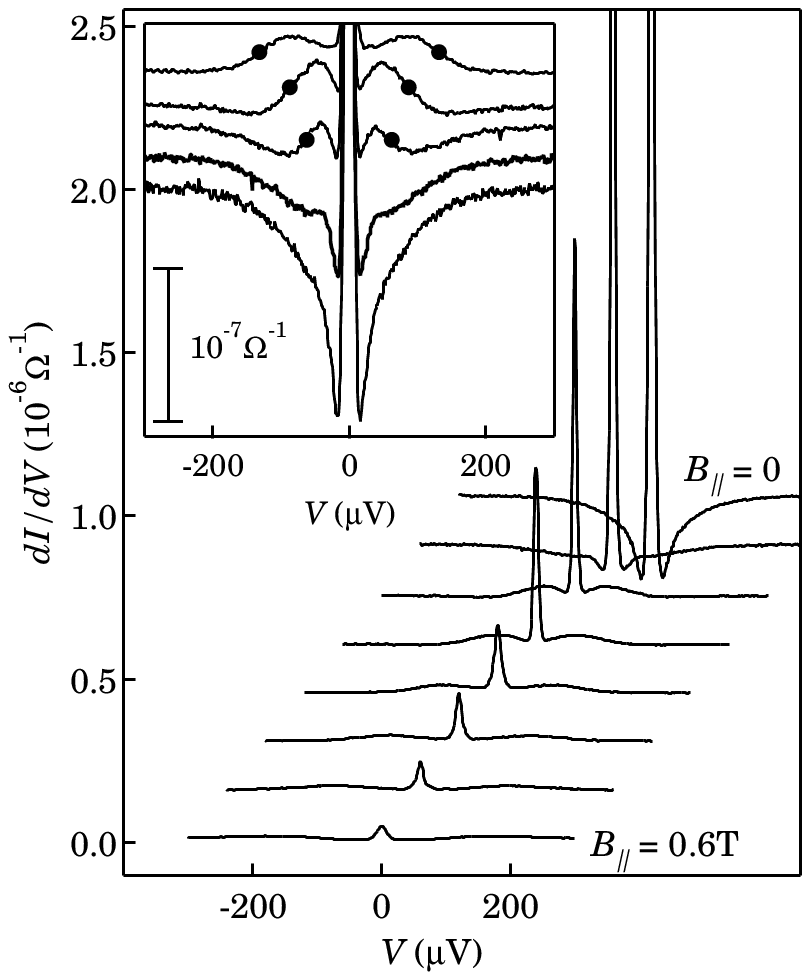} \epsfxsize=2.5in
\epsffile{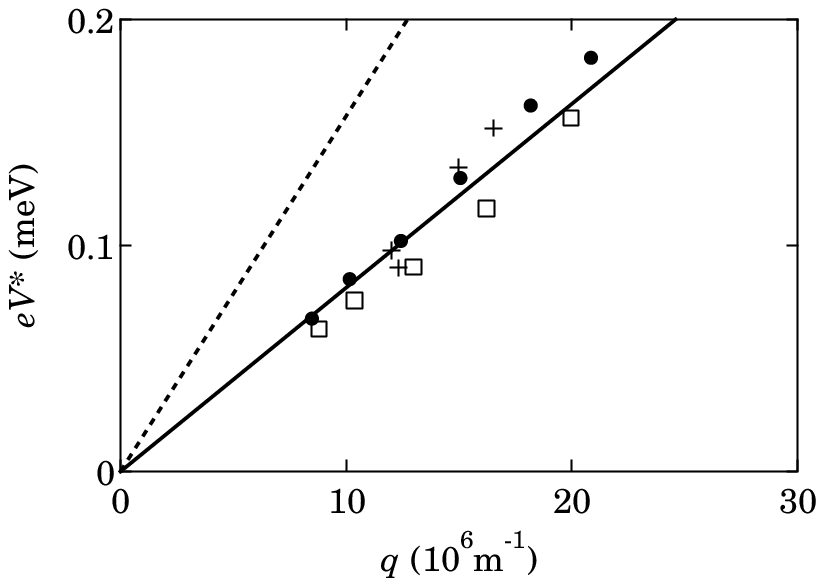} } \caption{ Left panel: Differential
conductance vs.~voltage for a variety of values of the parallel B
magnetic $B_\parallel$.  Inset: Magnified view showing the
Goldstone feature dispersing outward in voltage with increasing
$B_\parallel$.  Black dots indicate inflection points which are
used to determine the mode dispersion shown in the right panel.
The velocity agrees to within about a factor of two of the value
$\sqrt{\frac{\rho_s}{\Gamma}} \sim 10^4$m/s predicted from
Hartree-Fock estimates of the spin stiffness and the
compressibility parameters $\rho_s$ and $\Gamma$. After Spielman
et al. Ref.~\cite{jpegoldstone} }
\label{fig:tunnel2}
\end{figure}

Further evidence for the broken symmetry comes from observation of
the goldstone mode associated with the superfluidity.  By applying
a magnetic field in the plane of the 2DEGs, the tunnelling
electron picks up a finite in-plane momentum due to the Lorentz
force. This momentum $\hbar q$ goes into exciting a goldstone
boson with energy $\hbar\omega \sim \hbar c q$.  The energy
required to produce the boson should cause a small feature in the
tunnel I-V characteristic at bias voltage $eV=\hbar\omega$ which
shifts continuously with applied magnetic field.
\cite{ady2PRL,leonleoPRL,fogler} This goldstone mode feature has
been found by the Caltech group \cite{jpegoldstone}. The left
panel of Fig.~(\ref{fig:tunnel2}) shows that application of the
parallel magnetic field fairly quickly kills the central peak and
a small side feature appears which disperses outward with
increasing $B_\parallel$.  Spielman et al.~identify the inflection
point as the center of this derivative feature and plot the
resulting dispersion curve as shown in the right panel of
Fig.~(\ref{fig:tunnel2}). The dispersion is indeed linear and
agrees to within about a factor of two of the predicted mode
velocity of $\sim 10^4$m/s. It is perhaps not surprising that the
measured mode velocity is somewhat lower since quantum
fluctuations neglected in the Hartree-Fock approximation will
lower the spin stiffness.

\section{Open Issues}

At the present time, there are still a variety of open issues.
Experimentally there is now very strong evidence for the
interlayer phase coherent state and the corresponding
superfluidity in the antisymmetric channel.  The
Kosterlitz-Thouless phase transition seems to be occurring as
expected because the coherent tunnelling peak at zero bias appears
at temperature scales which are roughly consistent with the
predicted value of the pseudospin stiffness as well as the
observed and predicted goldstone mode velocity.  However no direct
measurement of the universal jump in superfluid stiffness has been
possible so far. An open theoretical issue is that we do not have
a complete microscopic understanding of the
dissipation/decoherence mechanism which gives a finite width and
height to the tunnelling peak.
\cite{ady2PRL,leonleoPRL,fogler,yogesh2} We do not understand why
the central peak is not destroyed more rapidly with the addition
of $B_\parallel$.  The peak is still visible even when the
Goldstone feature has moved out far enough to be distinct from it.
(Most likely this is due to disorder but a quantitative model is
lacking.) Finally, we do not have a good understanding of the
nature of the quantum phase transition or transitions that occur
as the layer spacing is increased. Various scenarios have been
suggested theoretically \cite{bonesteel,chetan,veillette} and
there is some numerical evidence hinting that there might be a
single weakly first order transition.  \cite{Schlieman}

\section*{Acknowledgments}
This work was supported by NSF DMR-0087133 and DMR-0196503,
 and represents
long-standing collaborations with many colleagues including Allan
MacDonald, Ady Stern, J. Schlieman, Ning Ma, K. Moon, and Kun
Yang. I also would like to thank J. P. Eisenstein and his group
for numerous helpful discussions of their experiments.

\section*{Dedication}  I would like to dedicate this paper to
the memory of William Caswell who was tragically killed on
September 11, 2001. Bill was a couple years ahead of me in
graduate school and gave me invaluable help in my effort to become
a physicist.

\end{document}